\documentclass[preprint,prb,showpacs]{revtex4}

\usepackage{graphicx}
\usepackage{dcolumn}
\usepackage{bm}
\usepackage{natbib}
\usepackage[english]{babel}
\usepackage{amsmath}
\usepackage{float}

\begin{document}

\title{\boldmath Correlation effects obtained from optical spectra of Fe-pnictides using an extended Drude-Lorentz model analysis\unboldmath}

\author{Seokbae Lee$^{1}$, Yu-Seong Seo$^{1}$, Seulki Roh$^1$, Dongjoon Song$^{2,3}$, Hirosh Eisaki$^2$, Jungseek Hwang$^{1}$}

\affiliation{$^1$Department of Physics, Sungkyunkwan University, Suwon, Gyeonggi-do 16419, Republic of Korea\\ $^2$National Institute of Advanced Industrial Science and Technology, Tsukuba 305-8568, Japan \\ $^3$Department of Physics and Astronomy, Seoul National University, Seoul 08826, Republic of Korea.}

\date{\today}

\begin{abstract}

We introduce an analysis model, an extended Drude–Lorentz model, and apply it to Fe-pnictide systems to extract their electron–boson spectral density functions (or correlation spectra). The extended Drude–Lorentz model consists of an extended Drude mode for describing correlated charge carriers and Lorentz modes for interband transitions. The extended Drude mode can be obtained by a reverse process starting from the electron–boson spectral density function and extending to the optical self-energy and, eventually, to the optical conductivity. Using the extended Drude–Lorentz model, we obtained the electron–boson spectral density functions of K-doped BaFe$_2$As$_2$ (Ba-122) at four different doping levels. We discuss the doping-dependent properties of the electron–boson spectral density function of K-doped Ba-122. We also can include pseudogap effects in the model using this approach. Therefore, this approach is very helpful for understanding and analyzing measured optical spectra of strongly correlation electron systems, including high-temperature superconductors (cuprates and Fe-pnictides).
\end{abstract}

\maketitle

\section{Introduction}

Fe-pnictide superconductors have been intensively studied since their discovery\cite{kamihara:2006}. They are known as multiband systems because they have multiple orbitals at the Fermi level\cite{lebegue:2007,ding:2008,kondo:2008,mazin:2008,subedi:2008}. Compared with single-band copper oxide superconductors, multiband systems may exhibit interesting features such as multiple superconducting gaps\cite{ding:2008,kondo:2008,wu:2010a,heumen:2010,dressel:2011,dai:2013} and non-trivial gap symmetries\cite{bang:2017}. Fe-pnictide systems are also known as correlated electron systems\cite{qazilbash:2009}. This is because the measured optical spectra reveal multiband superconducting gaps\cite{wu:2010a,dai:2013a} and correlations between electrons\cite{qazilbash:2009,yang:2009a,wu:2010,hwang:2015}. We note that the same analysis method, which has been used to extract information on correlations from measured optical spectra of cuprates\cite{hwang:2007}, has been approximately applied to extract the electron-boson spectral density function from measured optical spectra of Fe-pnictides\cite{yang:2009a,wu:2010,hwang:2015}. An analysis of the optical spectra of multiband Fe-pnictide systems in the normal state also has been performed using two (narrow coherent and broad incoherent) Drude modes\cite{tu:2010,wu:2010b,dai:2013,min:2013,lee:2015}. In the latter analysis, researchers used the two Drude modes to describe the charge carriers of a multiband system in the normal state. Further, there is another method for extracting the electron–boson spectral density function from the measured optical spectra of Fe-pnictide superconducting systems in the superconducting state by using two parallel transport channels\cite{hwang:2016}. In the case of the analysis with two Drude modes, the correlation effects might be implicitly included in the two Drude modes. In general, the correlation may divide the optical spectral weight of charge carriers into coherent and incoherent components\cite{hwang:2008a}. Therefore, the narrow Drude mode may contain most of the coherent components, while the broad Drude mode may contain most of the incoherent components. However, information on the correlations cannot be obtained explicitly from the aforenoted analysis with the two Drude modes. So far, we discuss low-energy correlations of Fe-pnictides. However, there are other intriguing topics of Fe-pnictides on correlations with higher energy scales, such as orbital selective correlations \cite{mazin:2010,nakamura:2010,qian:2011} and Hund rule couplings\cite{wang:2012,schafgans:2012}.

In this paper, we introduce a method to explicitly obtain the information on the low-energy correlations in correlated electron systems. First, we obtain {\it an extended Drude mode}, which can be defined based on the extended Drude model formalism\cite{puchkov:1996,hwang:2004}. This extended Drude mode can be employed to describe correlated charge carriers; therefore, it can be used to explicitly reveal correlation effects from the measured optical spectra of correlated electron systems, including Fe-pnictide systems. In this study, the extended Drude modes are obtained from input electron–boson spectral densities using an inverse process that has been previously developed and used for analyzing measured optical spectra\cite{hwang:2015a,hwang:2016}. To simulate measured optical conductivity, additional Lorentz modes are added to describe interband transitions generally situated in the high-energy region. Using this approach, we extract the electron–boson spectral density functions of K-doped BaFe$_2$As$_2$ (Ba-122) at various doping levels. We compare the results obtained using the extended Drude mode with those obtained using the two Drude modes. Furthermore, we obtain the doping-dependent mass renormalization factor (or correlation strength) from the extracted electron–boson spectral density function. Particularly, we observe a dome-shaped mass renormalization factor as a function of doping. This factor is different from that of cuprates, which monotonically increases as the doping decreases. We demonstrate that this difference can be associated with pseudogaps, which exist in Fe-pnictide superconducting systems\cite{xu:2011,moon:2012}.

\section{Analysis models}

We briefly describe the two models used in this study: One is a two-Drude (TD)–Lorentz (or TD-Lorentz) model, which consists of two Drude modes for describing intraband transitions (or itinerant charges) and Lorentz modes for interband transitions. In the TD-Lorentz model, optical conductivity can be described as
\begin{eqnarray}\label{eq1}
\tilde{\sigma}(\omega) &=& \tilde{\sigma}_{TD}(\omega) -\sum_{k}i\Big{(}\frac{\Omega_{k,p}^2}{4\pi}\Big{)}\frac{\omega}{\omega^2-\omega_k^2+i\omega\gamma_k},\nonumber \\ \tilde{\sigma}_{TD}(\omega) &=& \sum_{i = 1}^{2}i\Big{(} \frac{\Omega_{i,Dp}^2}{4 \pi} \Big{)} \frac{1}{\omega+i\tau_{i,imp}^{-1}},
\end{eqnarray}
where $\tilde{\sigma}(\omega)$ is the complex optical conductivity, $\tilde{\sigma}_{TD}(\omega)$ represents the two Drude (TD) modes, and $\Omega_{i,Dp}$ and $\tau_{i,imp}^{-1}$ are the plasma frequency and the impurity scattering rate of {\it i}th Drude mode ($i$ = 1 or 2), respectively. $\omega_k$, $\Omega_{k,p}^2$, and $\gamma_k$ are the resonance frequency, the strength, and the damping parameter of the {\it k}th Lorentz mode. In general, one of the Drude modes is called a narrow (or coherent) Drude and the other is termed a broad (or incoherent) one\cite{wu:2010,dai:2013}. This analysis model, which is legitimate, has been applied for various multiband systems\cite{tu:2010,wu:2010,dai:2013,lee:2015,homes:2015} and spawned many interesting findings including a hidden non-Fermi liquid behavior in Ba$_{0.6}$K$_{0.4}$Fe$_2$As$_2$\cite{dai:2013}.

The other analysis model is an extended Drude–Lorentz (ED-Lorentz) model. Here, we replace the two Drude modes ($\tilde{\sigma}_{TD}(\omega)$) with an extended Drude mode ($\tilde{\sigma}_{ED}(\omega)$). Therefore, the ED-Lorentz model consists of an extended Drude mode for describing the correlated charge carriers and Lorentz modes for interband transitions. The extended Drude mode can be obtained from an input electron–boson spectral density function ($I^2B(\omega)$) using a reverse process\cite{hwang:2015a}. Here $I$ is the coupling constant between an electron and a force-meditating boson and $B(\omega)$ is the boson spectrum. The reverse process consists of a series of steps starting from $I^2B(\omega)$, obtaining the optical conductivity, and eventually, to the reflectance spectrum\cite{hwang:2015a}. For a more detailed description of the reverse process, the readers can refer to Ref. \cite{hwang:2015a}. To get the optical conductivity of the extended Drude mode, we start from an input $I^2B(\omega)$ for the extended Drude mode, get the imaginary part of the optical self-energy of the extended Drude mode ($-2\Sigma^{op}_{ED,2}(\omega)$) or the optical scattering rate of the extended Drude mode ($1/\tau^{op}_{ED}(\omega)$) using the generalized Allen formula\cite{allen:1971,shulga:1991}, which can be described as $-2\Sigma^{op}_{ED,2}(\omega) \equiv 1/\tau^{op}_{ED}(\omega) = (\pi/\omega) \int_0^{\infty}d\Omega \: I^2B(\Omega)\{ 2\omega \coth(\Omega/2T)-(\omega+\Omega)\coth[(\omega+\Omega)/2T]+(\omega-\Omega)\coth[(\omega-\Omega)/2T] \}$, where $T$ is the absolute temperature. Then, we calculate the real part of the optical self-energy of the extended Drude mode ($-2\Sigma^{op}_{ED,1}(\omega)$) using the Kramers-Kronig relation as the real and imaginary parts of the self-energy form a Kramers-Kronig pair\cite{hwang:2015a,hwang:2020}, which can be written as $-2\Sigma^{op}_{ED,1}(\omega) =-(2/\pi)\int_0^{\infty}d\omega'\:\omega'[-2\Sigma^{op}_{ED,2}(\omega')]/(\omega'^2-\omega^2)$. Eventually, we obtain the complex optical conductivity of extended Drude mode using the extended Drude model formalism\cite{hwang:2004}, which can be written as
\begin{equation}\label{eq2}
\tilde{\sigma}_{ED}(\omega) \equiv i\Big{(} \frac{\Omega_{p,ED}^2}{4 \pi} \Big{)} \frac{1}{\omega+[-2\tilde{\Sigma}^{op}_{ED}(\omega)]},
\end{equation}
where $\tilde{\sigma}_{ED}(\omega)$ is the extended Drude mode and $-2\tilde{\Sigma}^{op}_{ED}(\omega)$ is the corresponding complex optical self-energy to the extended Drude mode. $\Omega_{p,ED}^2/8$ is the total spectral weight of the charge carriers (or ED mode) in a correlated material system, where $\Omega_{p,ED}$ is the plasma frequency of the extended Drude mode. Therefore, the measured optical conductivity can be described by the following model:
\begin{equation}\label{eq3}
\tilde{\sigma}(\omega) = \tilde{\sigma}_{ED}(\omega)-\sum_{k}i\Big{(}\frac{\Omega_{k,p}^2}{4\pi}\Big{)}\frac{\omega}{\omega^2-\omega_k^2+i\omega\gamma_k}.
\end{equation}
It is worth noting that the optical self-energy ($-2\tilde{\Sigma}^{op}(\omega)$) corresponding to the total optical conductivity ($\tilde{\sigma}(\omega)$) can be defined, based on the extended Drude model formalism, as $-2\tilde{\Sigma}^{op}(\omega)\equiv i\frac{\Omega_p^2}{4\pi}\frac{1}{\tilde{\sigma}(\omega)}-\omega$. In general, the real and imaginary part of the total optical self-energy ($-2\tilde{\Sigma}^{op}(\omega)$) do not form a Kramers-Kronig pair. However, the real and imaginary parts of the optical self-energy of the ED mode ($-2\tilde{\Sigma}^{op}_{ED}(\omega)$) self-consistently form a Kramers-Kronig pair\cite{hwang:2015a,hwang:2020}. We will discuss this further in the discussion section.

\section{Results and discussions}

\begin{figure}[!htbp]
  \vspace*{-0.5 cm}%
  \centerline{\includegraphics[width=4.0 in]{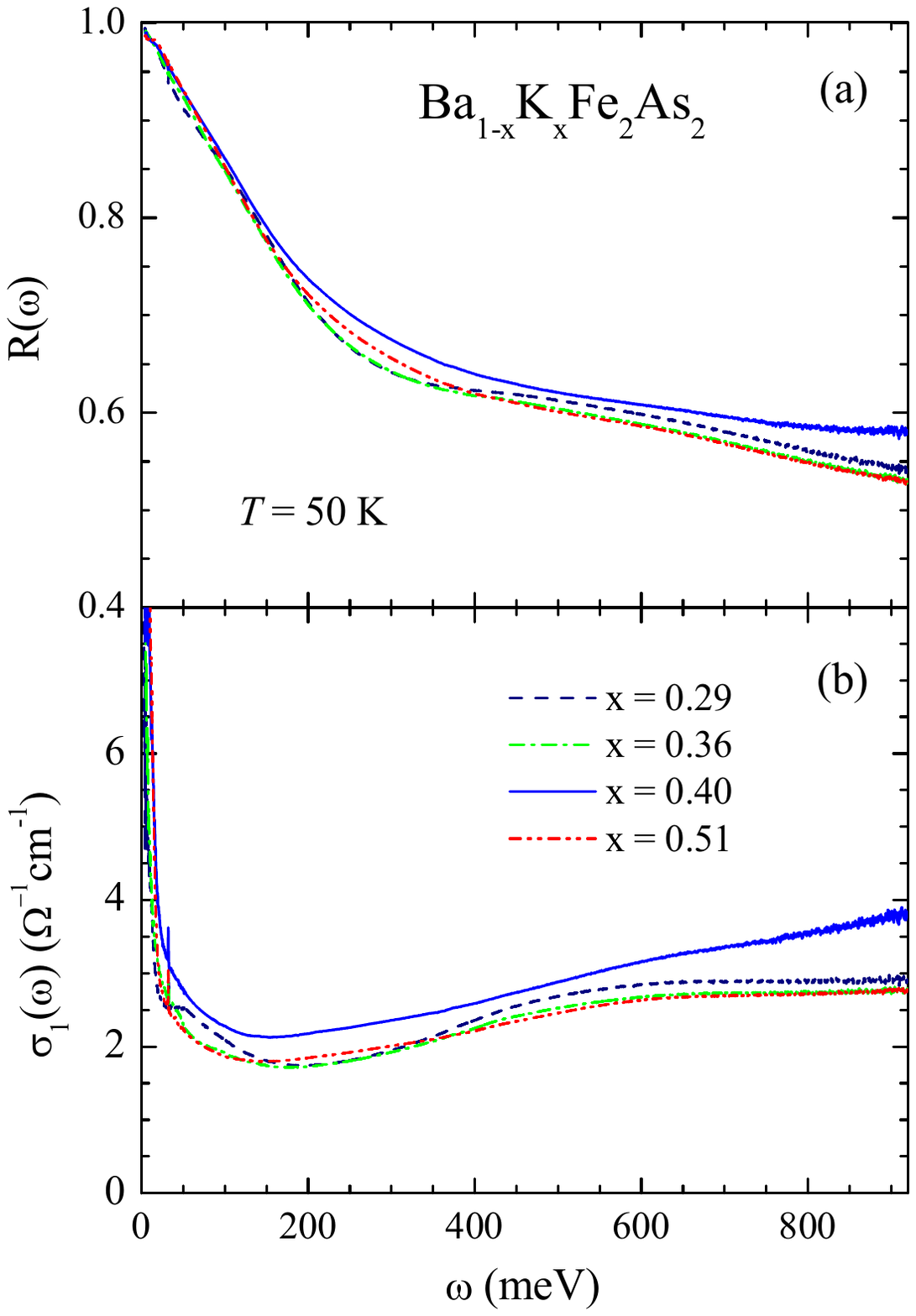}}%
  \vspace*{-1.3 cm}%
\caption{(Color online) Measured reflectance and optical conductivity spectra at $T =$ 50 K. (a) Measured reflectance spectra of four K-doped Ba-122 single crystal samples. (b) The corresponding optical conductivity spectra obtained using the Kramers-Kronig analysis.}
 \label{fig1}
\end{figure}

We investigate K-doped Ba-122 (Ba$_{1-x}$K$_x$Fe$_2$As$_2$) single crystals at four different doping levels namely $x =$ 0.29, 0.36, 0.40, and 0.51, which have the superconducting transition temperatures ($T_c$) of 35.9 K, 38.5 K, 38.5 K, and 34.0 K, respectively. The sample at $x =$ 0.40 is optimally doped. The K-doped single crystal samples are grown using a self-flux technique\cite{nakajima:2014}. We take reflectance spectra (35 - 8000 cm$^{-1}$ or $\sim$4 meV - 1.0 eV) of the four single crystal samples at various temperatures using an {\it in-situ} gold evaporation technique\cite{homes:1993} and a continuous liquid helium flow cryostat. In this study, we focus on the measured optical spectra in the normal state ($T$ = 50 K). We obtained the optical conductivity spectra from the measured reflectance spectra using a Kramers-Kronig (KK) analysis\cite{wooten,tanner:2019}. For the KK analysis, we take extrapolations to zero frequency and infinity. For the extrapolation to zero frequency, we use the Hagen-Rubens relation ($1-R(\omega) \propto \sqrt{\omega}$). For the extrapolation to infinity, we use an available reported data\cite{dai:2013a} up to 40,000 cm$^{-1}$, $R(\omega) \propto \omega^{-1}$ from 40,000 to 10$^6$ cm$^{-1}$, and the free electron response ($R(\omega) \propto \omega^{-4}$) above 10$^6$ cm$^{-1}$.  We show the measured reflectance spectra and the corresponding optical conductivities in Fig. \ref{fig1}.

\begin{figure}[!htbp]
  \vspace*{-0.3 cm}%
  \centerline{\includegraphics[width=4.5 in]{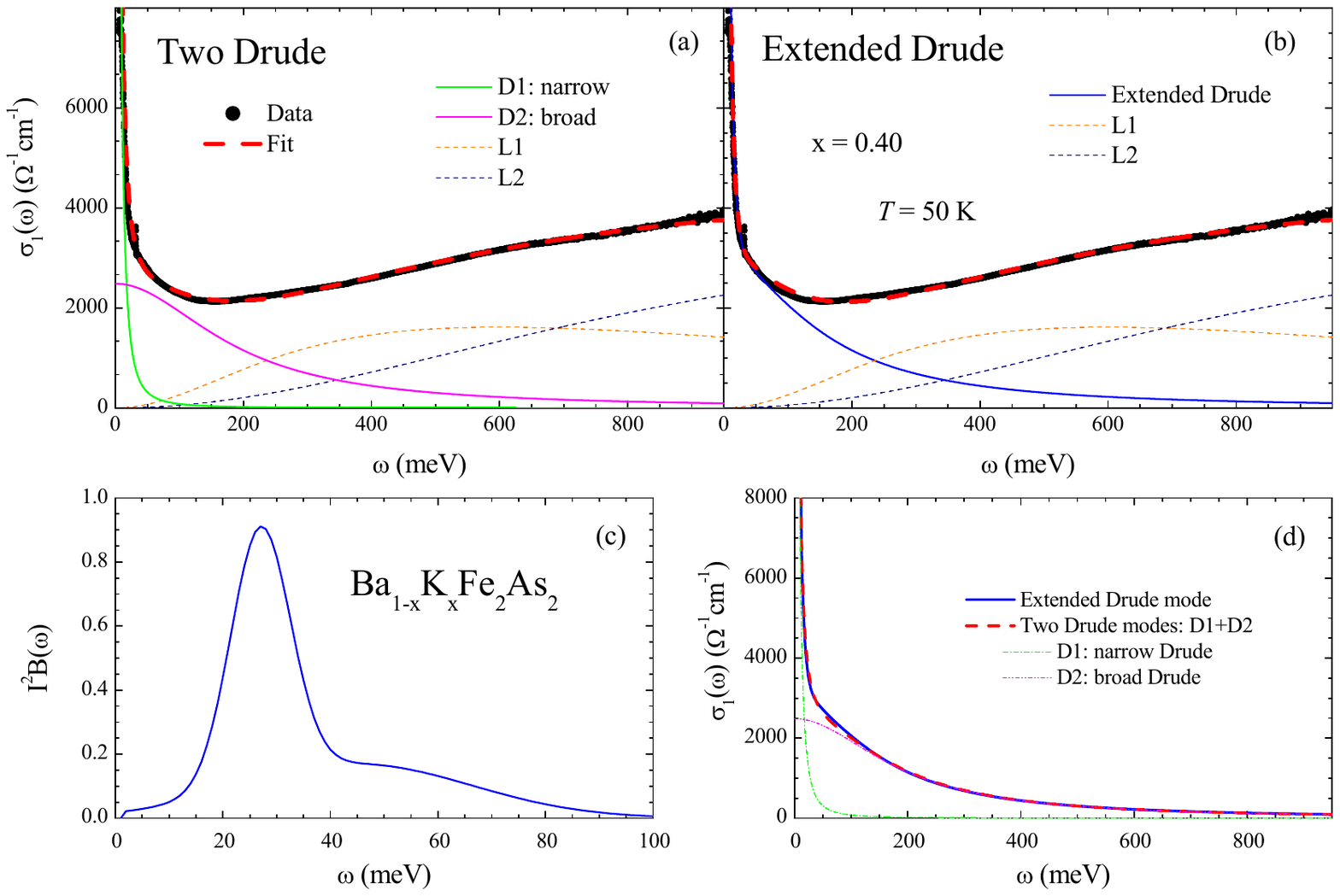}}%
  \vspace*{-0.8 cm}%
\caption{(Color online) Comparison of results obtained via both the TD-Lorentz and the ED-Lorentz model analyses. (a) Data and fit obtained by using the TD-Lorentz model. (b) Data and fit obtained by using the ED-Lorentz model. (c) The electron–boson spectral density function obtained by the ED-Lorentz model analysis. (d) Comparison of the optical conductivity of the extended Drude mode with the combined optical conductivity of the two Drude modes.}
 \label{fig2}
\end{figure}

We use the two (TD-Lorentz and ED-Lorentz) models to analyze the optical conductivity up to 950 meV. In Fig. \ref{fig2}, we compare the results obtained by applying the two models to the optical conductivity (at 50 K) of the optimally K-doped Ba-122. Fig. \ref{fig2}(a) and \ref{fig2}(b) show the data and fits obtained using the TD-Lorentz and ED-Lorentz models, respectively, below 950 meV. We also separately show the two Drude modes, extended Drude mode, and Lorentz modes. We used the same Lorentz modes for both the fittings. The overall quality of fits was similar for the two models. Fig. \ref{fig2}(c) shows the electron-boson spectral density function ($I^2B(\omega)$) obtained using the ED-Lorentz model. Here, we use a model $I^2B(\omega)$ that consists of two Gaussian functions: a sharp Gaussian function and a broad Gaussian function. It is worth noting that we adopted this model $I^2B(\omega)$ from earlier studies of Fe-pnictides and cuprates\cite{hwang:2006,carbotte:2011,hwang:2016}. We used the reverse process to obtain the extended Drude mode from this input electron-boson spectral density function\cite{hwang:2015a}. There are six fitting parameters for $I^2B(\omega)$, which consists of two Gaussian functions, and one for $\Omega_{p,ED}$. By adjusting the seven fitting parameters, we obtained a reasonable fit in the low-frequency region, as shown in \ref{fig2}(b). The extracted complex optical self-energies of the four samples are shown in Supplementary Materials. In Fig. \ref{fig2}(d), we show the optical conductivities of both the Drude modes and the ED Drude mode; the two optical conductivities agree well. This result indicates that, in normal state, one may not need to consider two separate transport channels to describe charge carriers in the multiband systems. We speculate that the band characteristics are fuzzed by thermal excitations and/or correlations. However, to expose a hidden non-Fermi-liquid behavior, the two-Drude approach is useful, as reported in the literature\cite{dai:2013,lee:2015}. For superconducting state, a two-channel approach must be used because the multiple bands provide superconducting gaps of different sizes, as reported in the literature\cite{ding:2008,hwang:2016}.

The correlation effects can be estimated from the obtained $I^2B(\omega)$. The mass renormalization factor ($\lambda$) caused by the correlations can be estimated from $I^2B(\omega)$ using the well-known formula, $\lambda \equiv 2\int^{\omega_c}_0[I^2B(\omega)/\omega] \:d\omega$, where $\omega_c$ is the cutoff frequency. In our case, $\omega_c$ is 100 meV because $I^2B(\omega)$ is negligibly small above 100 meV. The estimated mass renormalization factor of this sample is 1.41. From this quantity, one can obtain separate spectral weights for the coherent and incoherent components\cite{hwang:2008a}. The coherent and incoherent components correspond to $1/(1+\lambda)$ and $\lambda/(1+\lambda)$ fractions of the total spectral weight of charge carriers, respectively. Specifically, the coherent and incoherent components account for 41\% and 59\% of the total spectral weight of the charge carriers for this optimally doped sample ($x$ = 0.40), respectively.

\begin{figure}[!htbp]
  \vspace*{-0.3 cm}%
  \centerline{\includegraphics[width=4.5 in]{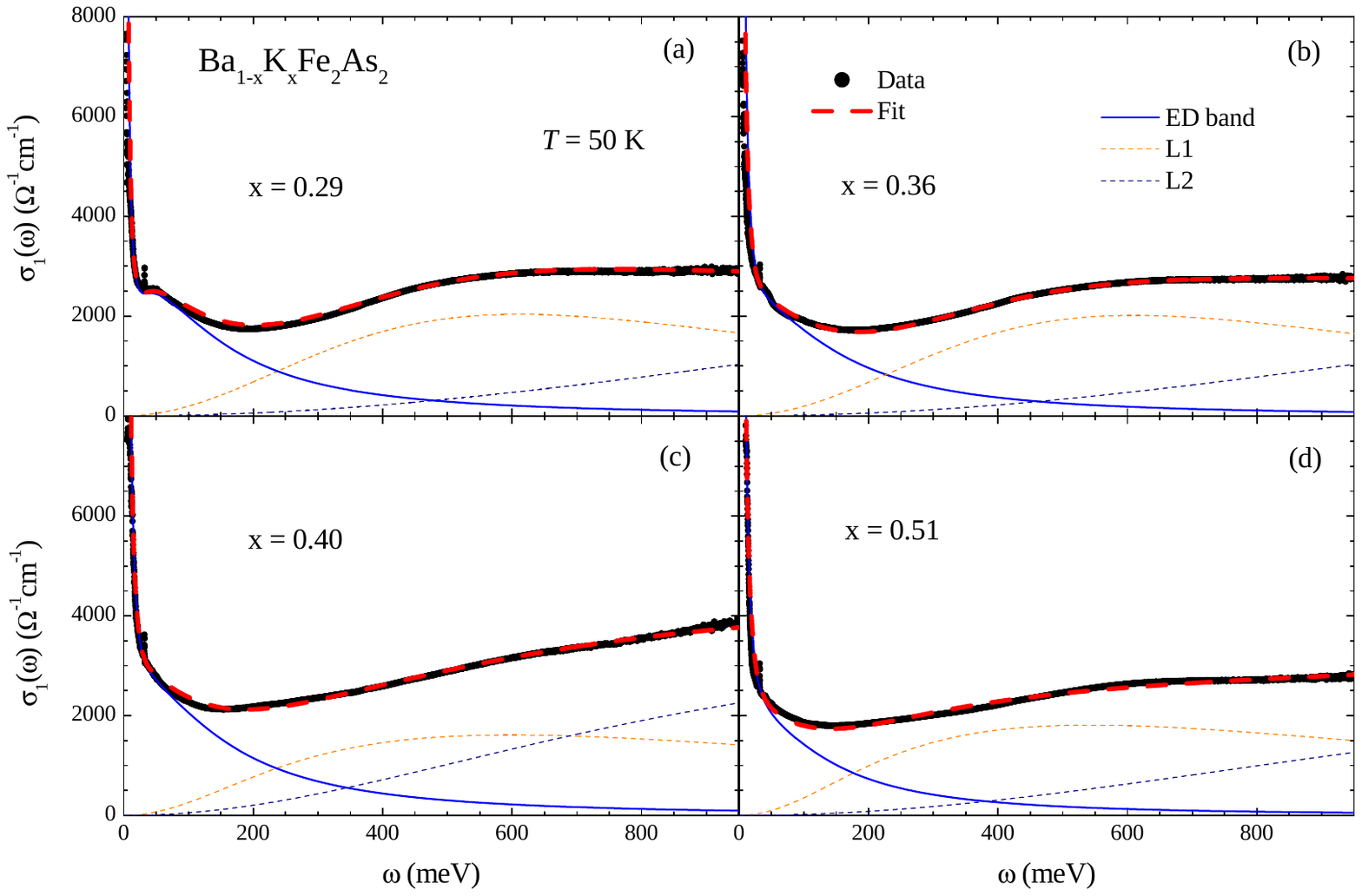}}%
  \vspace*{-0.8 cm}%
\caption{(Color online) Data and ED-Lorentz model fits of four K-doped Ba-122 (Ba$_{1-x}$K$_x$Fe$_2$As$_2$) samples at 50 K: (a) $x =$ 0.29, (b) $x =$ 0.36, (c) $x =$ 0.40, and (d) $x =$ 0.51.}
 \label{fig3}
\end{figure}

\begin{figure}[!htbp]
  \vspace*{-0.3 cm}%
  \centerline{\includegraphics[width=4 in]{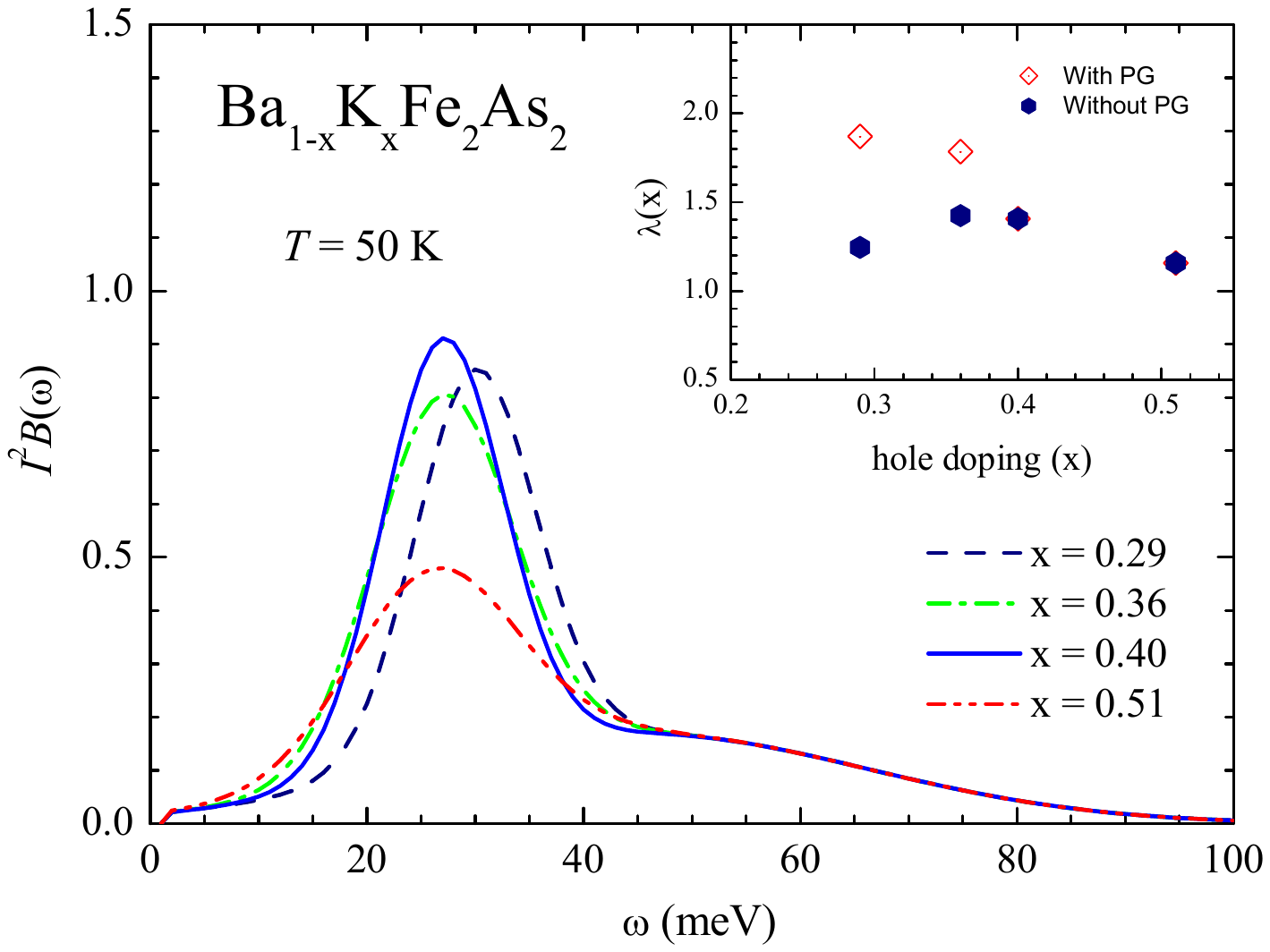}}%
  \vspace*{-0.8 cm}%
\caption{(Color online) The extracted electron-boson spectral density function, $I^2B(\omega)$ of our four samples. In the inset, the mass renormalization factors of the two cases with (open diamond) and without (solid hexagon) including pseudogaps (PG) are shown.}
 \label{fig4}
\end{figure}

\begin{figure}[!htbp]
  \vspace*{-0.3 cm}%
  \centerline{\includegraphics[width=4.5 in]{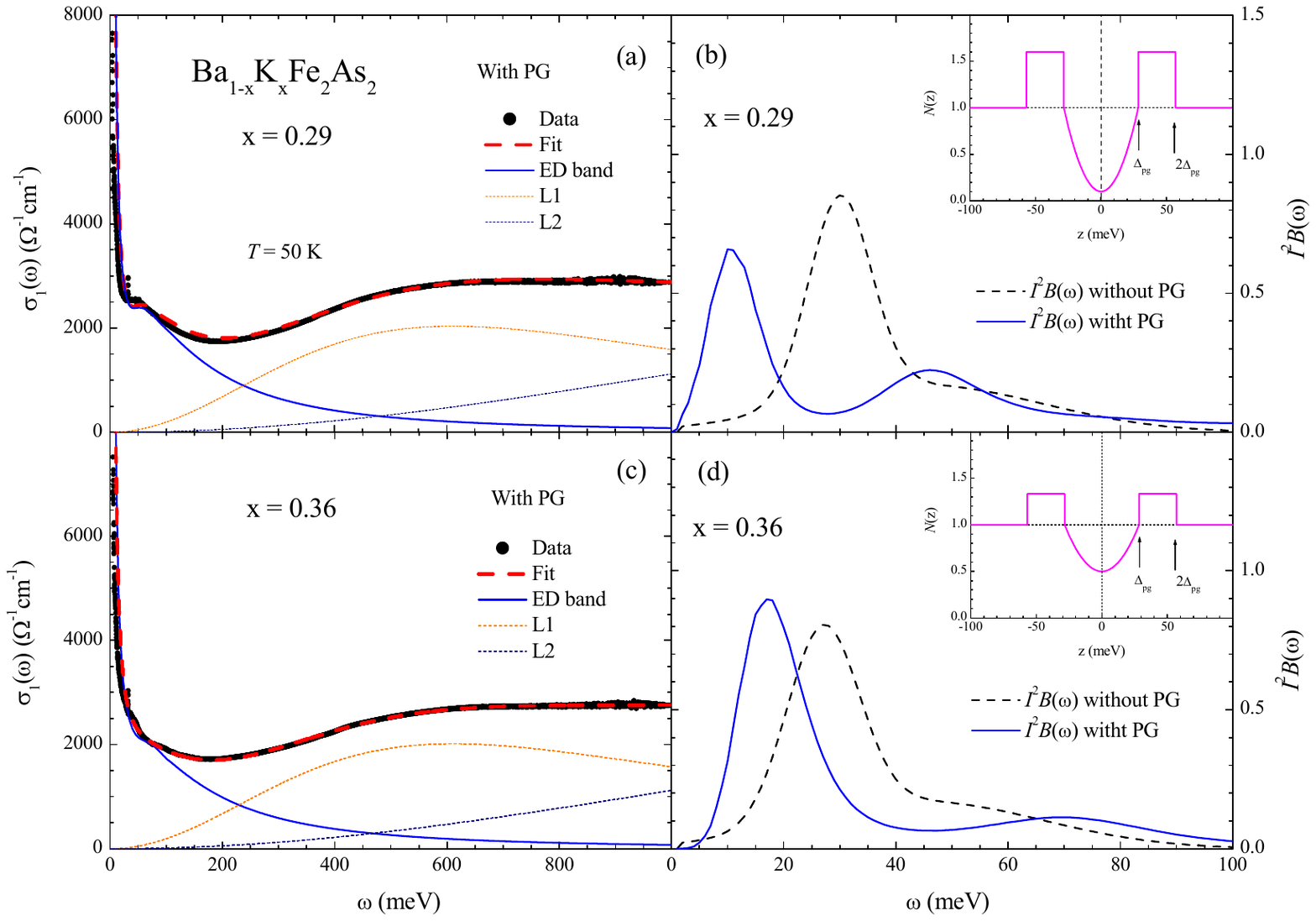}}%
  \vspace*{-0.8 cm}%
\caption{(Color online) Data and ED-Lorentz model fits with including pseudogaps. (a) and (c) Data and fits of the two underdoped samples with $x=$ 0.29 and 0.36. (b) and (d) The extracted electron-boson spectral density function, $I^2B(\omega)$ of the two underdoped samples with pseudogaps (solid lines) and without pseudogaps (dashed lines). In the insets, the model pseudogaps are shown (see in the text).}
 \label{fig5}
\end{figure}

We apply the ED-Lorentz model to the two underdoped ($x$ = 0.29 and 0.36) and one overdoped ($x$ = 0.51) samples. We also used two (sharp and broad) Gaussian functions for $I^2B(\omega)$. The underdoped samples may have pseudogaps\cite{xu:2011,moon:2012,wang:2012,marsik:2013}; nevertheless, for now, we do not include pseudogap in our model. In a later paragraph, we include the pseudogaps in the analysis to show pseudogap effects on the resulting $I^2B(\omega)$ (see Fig. \ref{fig5}). In Fig. \ref{fig3}, we show data and fits of four K-doped Ba-122 samples, including the optimally doped sample ($x =$ 0.40), at 50 K. The overall shapes of the real parts of optical conductivities of the four samples are similar: a sharp increase below 50 meV, a broad shoulder below 200 meV, and a broad absorption peak near 500 meV. These are the generic features of the optical conductivity of Fe-pnictides\cite{hu:2009,kim:2010,dai:2013,min:2013}. The quality of fits was similar for all four samples. The ED mode and two Lorentz modes are displayed separately for each sample. The plasma frequencies of the extended Drude modes are 18800, 17200, 19200, and 16300 cm$^{-1}$ for $x =$ 0.29, 0.36, 0.40 and 0.51, respectively. We observed slight doping-dependent evolutions of the ED mode and the Lorentz modes. From these fittings, we obtained the electron-boson spectral density functions ($I^2B(\omega)$) of the four samples. For the fittings, we did not need to adjust the broad Gaussian peak in $I^2B(\omega)$ for all four samples. Therefore, we used the same broad Gaussian peak for all samples, as shown in Fig. \ref{fig4}.

In Fig. \ref{fig4}, we show the obtained electron-boson spectral density functions ($I^2B(\omega)$) of our four samples. For Bi$_2$Sr$_2$CaCu$_2$O$_{8+\delta}$ (Bi-2212), the energy of the sharp peak is proportional to the superconducting transition temperature ($T_c$)\cite{yang:2009}, and its intensity decreases rapidly in the overdoped region\cite{hwang:2007}. The obtained $I^2B(\omega)$ of K-doped Ba-122 also exhibits some doping dependencies. The peak intensities of underdoped and optimally doped samples are similar, but that of the overdoped sample is significantly reduced, similar to the case of Bi-2212; however, the peak position does not show a systematic change with doping. In the inset, we show the mass renormalization factors ($\lambda$) of the four samples at different K-doping levels. The mass renormalization factor (solid hexagon) shows a dome shape, which is different from that of Bi-2212; the $\lambda$ of Bi-2212 monotonically decreases as the doping increases\cite{hwang:2007a}. This dome-shaped renormalization factor looks similar to a peaked London penetration depth observed around the optimal doping in P-doped Ba-122\cite{hashimoto:2012}. However, we expect that these different doping-dependent behaviors of $\lambda$ of two material systems might be related to the pseudogaps. As we mentioned earlier, Fe-pnictide systems have been known to contain the pseudogaps\cite{shimojima:2014,moon:2014}. The pseudogaps exhibit similar temperature- and doping-dependent behaviors of those in cuprates. So far, in our analysis model, we did not include the pseudogaps. In general, if pseudogaps are included, the $I^2B(\omega)$ spectrum, including the sharp peak, will be shifted to a lower energy\cite{hwang:2012}, resulting in an increase in the mass renormalization factor.

To demonstrate the pseudogap effect on the electron-boson spectral density function, we include the pseudogap (PG) in the analysis for the two underdoped samples ($x =$ 0.29 and 0.36). In this case, the optical scattering rate of the extended Drude mode can be written in a more generalized form\cite{sharapov:2005} as $ 1/\tau^{op}_{ED}(\omega) \equiv -2\Sigma^{op}_{ED,2}(\omega)= (\pi/\omega) \int_0^{\infty}d\Omega \: I^2B(\Omega)\int_{-\infty}^{+\infty} dz[N(z-\Omega)+N(-z+\Omega)][n_B(\Omega)+f(\Omega-z)] [f(z-\omega)-f(z+\omega)]$, where $n_B(\omega)$ and $f(\omega)$ are the Bose-Einstein and Fermi-Dirac distribution functions, respectively, and $N(z)$ is the normalized density of states, which can be used to describe the pseudogap. The pseudogap is modelled\cite{hwang:2008b,hwang:2012} as $N(z) = N_0+(1-N_0)(z/\Delta_{pg})^2 \:\:\mbox{for}\:\: |z|<\Delta_{pg}, (5-2N_0)/3 \:\:\mbox{for}\:\: |z|\in(\Delta_{pg}, 2\Delta_{pg}),\:\: \mbox{and} \:\: 1 \:\:\mbox{for}\:\: |z|\geq 2\Delta_{pg}$, where $N_0$ is the normalized density of states at the Fermi level and $\Delta_{pg}$ is the size of the pseudogap. This pseudogap model has been introduced and used for analyzing optical spectra of underdoped cuprate systems\cite{hwang:2008,hwang:2008b,hwang:2012,hwang:2013}. We note that, in this PG model, the density of states loss in the pseudogap is recovered just above the pseudogap\cite{hwang:2008,hwang:2008b} as shown in the insets of Fig. \ref{fig5}. We assume that the same size of pseudogap ($\Delta_{pg}$) is $\sim$28.4 meV for the two samples and the depths ($1-N_0$) of the pseudogap are 0.9 and 0.5 for $x =$ 0.29 and 0.36, respectively. Fig. \ref{fig5}(a) and (c) show the data and fits of the two underdoped samples. We used the same plasma frequencies of the extended Drude modes as we used for the cases of previous fittings without including the PG. The new electron-boson spectral density functions, $I^2B(\omega)$, obtained including the PG are shown in Fig. \ref{fig5}(b) and (d). We clearly observe the sharp Gaussian peak shifts to a lower energy. The amount of the shift increases as the strength (or depth) of pseudogap increases as reported in a published paper\cite{hwang:2012}. The new renormalization factors are 1.87 and 1.78 for $x =$ 0.29 and 0.36, respectively, as shown in the inset of Fig. \ref{fig4}. The new renormalization factor monotonically increases as the doping decreases, which is a similar doping-dependent behavior of the renormalization factor of hole-doped cuprates\cite{hwang:2007a}. In this study, we mainly focused on the introduction of the extended Drude-Lorentz model approach. Furthermore, we also included the pseudogaps in the analyses of underdoped samples and obtained more reasonable results; the new doping-dependent $\lambda$ was similar to that of cuprates, which one might expect considering the similar phase diagrams of the two material systems (Fe-pnictides and cuprates) and antiferromagnetic fluctuations as force-mediating bosons. However, there might be some other issues regarding the pseudogap for Fe-pnictides, such as the depth (or strength) and energy scale of the pseudogap.

The ED-Lorentz model, when considered exclusively, is not new because the Drude mode is simply replaced with the extended Drude mode in the prevalent Drude-Lorentz model. However, this approach has not been applied thus far to analyze multiband Fe-pnictide superconducting systems and single-band cuprates. In this regard, the proposed ED-Lorentz model is a intriguing and effective method for analyzing the optical spectra of correlated electron systems, including high-temperature superconductors. Herein, we describe {\it the extended Drude mode} more in detail: Similar to the case of the simple Drude mode, the real and imaginary parts of the extended Drude mode form a Kramers-Kronig pair. In the case of an ideal system that can be described with an extended Drude mode, the real and imaginary parts of the corresponding optical self-energy form a Kramers-Kronig pair as well\cite{hwang:2015a,hwang:2020}. Moreover, if we include additional Lorentz modes to realize a real correlated electron system, which exhibits both intraband and interband (optical) transitions, then the real and imaginary parts of the total (ED plus Lorentzian) optical conductivity still form a Kramers-Kronig pair holding the causality condition\cite{wooten}. However, in general, the real and imaginary parts of the corresponding total optical self-energy can no longer form a Kramers-Kronig pair; this is because the optical self-energy with multiple components is related to the optical conductivity with multiple components in the extended Drude formalism. In this sense, optical-self energy is not a completely well-defined optical quantity for describing the measured optical spectra. However, if only extended Drude mode can be extended by excluding all Lorentz modes from the measured optical spectra, the optical self-energy of the remaining extended Drude mode can be a well-defined optical quantity. Interestingly, in the cuprate systems, the extended Drude mode is relatively well isolated in a low-energy region because all of the Lorentz modes of the system are located in the high-energy region above $\sim$ 2 eV\cite{hwang:2007a}. By contrast, in Fe-pnictide systems, the extended Drude mode significantly overlaps with Lorentz modes located in the low-energy region\cite{benfatto:2011} (also see Fig. \ref{fig2}(b) and Fig. \ref{fig3}). Therefore, the optical self-energy is relatively better defined in cuprate systems than in Fe-pnictides. We note that both systems can be reasonably well analyzed to reveal the correlation effects by using the extended Drude-Lorentz model. It is worthwhile to note that, in our analysis, we assumed that no interband transitions exist below 200 meV other than tails of the Lorentz modes located at the higher energy. If there is a sharp and strong Lorentz mode below 200 meV we expect to be able to observe it because its line shape is different from that of the extended Drude mode\cite{benfatto:2011,hwang:2019}.

\section{Conclusions}

We introduced an interesting approach for analyzing the optical spectra of correlated electron systems. The approach was named "the extended Drude-Lorentz model". The extended Drude mode can be obtained from the electron-boson spectral density function using a reverse process\cite{hwang:2015a}. We compared the extended Drude-Lorentz model with the two-Drude-Lorentz model. The extended Drude-Lorentz model explicitly provides information on the correlations between charge carriers. We applied this approach to the measured optical conductivity spectra of K-doped Ba-122 single crystal samples at various doping levels in a wide (from underdoped to overdoped) doping region. We obtained the electron-boson spectral density functions ($I^2B(\omega)$) at the various doping levels. We also obtained doping-dependent mass renormalization factor ($\lambda$), which exhibits a dome shape. This factor is maximized near optimally doping level. The different doping-dependent behaviors of $I^2B(\omega)$ of the two high-temperature superconducting systems (cuprates and Fe-pnictides) might be associated with pseudogaps; we demonstrated that if we include the pseudogap in our model the doping-dependent mass renormalization factors of the two material systems became similar to each other. This approach will be helpful for conceptually understanding of measured optical spectra in the extended Drude-Lorentz model and also useful for analyzing measured optical spectra of strongly correlated electron systems, including high-temperature superconductors and heavy fermion systems.

%
%
\acknowledgments J.H. acknowledges financial support from the National Research Foundation of Korea (NRFK Grant No. 2019R1A6A1007307912, 2020R1A4A4078780, and 2021R1A2C101109811).

\bibliographystyle{apsrev4-1}
\bibliography{bib}

\end{document}